\documentclass[sigconf]{acmart}
 
\AtBeginDocument{%
\providecommand\BibTeX{{%
\normalfont B\kern-0.5em{\scshape i\kern-0.25em b}\kern-0.8em\TeX}}}

\copyrightyear{2021} 
\acmYear{2021} 
\setcopyright{acmcopyright}\acmConference[UCC'21]{2021 IEEE/ACM 14th International Conference on Utility and Cloud Computing }{December 6--9, 2021}{Leicester, United Kingdom}
\acmBooktitle{2021 IEEE/ACM 14th International Conference on Utility and Cloud Computing (UCC'21), December 6--9, 2021, Leicester, United Kingdom}
\acmPrice{15.00}
\acmDOI{10.1145/3468737.3494094}
\acmISBN{978-1-4503-8564-0/21/12}
\acmConference[UCC 2021]{14th International Conference on Utility and Cloud Computing}{December 06--09, 2021}{Leicester, UK}

\usepackage{graphicx}
\usepackage{color}
\usepackage{tabularx}
\usepackage{listings}
\usepackage{array}
\usepackage{url}
\usepackage{siunitx}
\usepackage{gensymb}
\usepackage{acronym}
\usepackage{xcolor, colortbl}
\usepackage{todonotes}
\usepackage{balance}
\usepackage{amsmath}
\usepackage{subcaption} 
\usepackage{braket}
\usepackage{longtable}
\usepackage{svg}

\usepackage{booktabs}

\newacro{TP}[TP]{True-Positive}
\newacro{TN}[TN]{True-Negative}
\newacro{FP}[FP]{False-Positive}
\newacro{FN}[FN]{False-Negative}
\newacro{TCM}[TCM]{Tool Condition Monitoring}
\newacro{ML}[ML]{Machine Learning}
\newacro{DDV}[DDVN]{Distributed Data Validation Network}
\newacro{DNN}[DNN]{Deep Neural Network}
\newacro{IIOT}[IIoT]{Industrial Internet of Things}

\definecolor{normal}{RGB}{213,229,255}
\definecolor{anomaly}{RGB}{232, 172, 140}

\usepackage{tikz, pgfplots}
\usetikzlibrary{calc, shapes, backgrounds, arrows.meta, positioning}
\tikzset{
    arrow_default/.style = {-{Latex[length=3mm]}},
    rectangle_empty/.style = {fill=none, rectangle},
    rectangle_none/.style = {fill=none, rectangle, draw=none},
    rectangle_none_vertical/.style = {fill=none, rectangle, rotate around={90:(-1,0.5)}, inner sep=10},
    rectangle_blue_vertical/.style = {fill=blue!20, rectangle, rotate around={90:(-1,0.5)}, inner sep=10},
    rectangle_red/.style = {fill=red!20, rectangle, inner sep=10},
    rectangle_gray/.style = {fill=gray!20, rectangle, inner sep=10},
    rectangle_blue/.style = {fill=blue!20, rectangle, inner sep=10},
    rectangle_orange/.style = {fill=orange!20, rectangle, inner sep=10},
    rectangle_green/.style = {fill=green!20, rectangle, inner sep=10},
    rectangle_orange_vertical/.style = {fill=orange!20, rectangle, rotate around={90:(-1,0.5)}, inner sep=10},
    rectangle_gray_vertical/.style = {fill=gray!20, rectangle, rotate around={90:(-1,0.5)}, inner sep=10},
    circle_gray/.style = {fill=gray!20, circle, inner sep=7},
    circle_gray_min/.style = {fill=gray!20, circle, inner sep=3},
    doc/.style={align=center, color=black, fill=white, shape=document, minimum width=10mm, minimum height=14mm, inner sep=2ex}
}

\definecolor{color1a}{RGB}{102,194,165} 
\definecolor{color2a}{RGB}{252,141,98} 
\definecolor{color3a}{RGB}{141,160,203} 
\definecolor{color4a}{RGB}{231,138,195} 
\definecolor{color5a}{RGB}{166,216,84}
\definecolor{color6a}{RGB}{255,217,47}
\definecolor{color7a}{RGB}{229,196,148}

\renewcommand\footnotetextcopyrightpermission[1]{}
\setcopyright{none}

\begin{document}

\title{QUDOS: Quorum-Based Cloud-Edge Distributed DNNs for Security Enhanced Industry 4.0}

\author{Kevin Wallis}
\affiliation{
\institution{University of Applied Sciences Furtwangen}
\country{Germany}
}
\email{kevin.wallis@hs-furtwangen.de}

\author{Christoph Reich}
\affiliation{
\institution{University of Applied Sciences Furtwangen}
\country{Germany}
}
\email{christoph.reich@hs-furtwangen.de}

\author{Blesson Varghese}
\affiliation{
\institution{Queen’s University Belfast}
\country{United Kingdom}
}
\email{b.varghese@qub.ac.uk}

\author{Christian Schindelhauer}
\affiliation{
\institution{University of Freiburg}
\country{Germany}
}
\email{schindel@tf.uni-freiburg.de}

\begin{abstract}
Distributed machine learning algorithms that employ Deep Neural Networks (DNNs) are widely used in Industry 4.0 applications, such as smart manufacturing. The layers of a DNN can be mapped onto different nodes located in the cloud, edge and shop floor for preserving privacy. The quality of the data that is fed into and processed through the DNN is of utmost importance for critical tasks, such as inspection and quality control. Distributed Data Validation Networks (DDVNs) are used to validate the quality of the data. However, they are prone to single points of failure when an attack occurs. This paper proposes \texttt{QUDOS}, an approach that enhances the security of a distributed DNN that is supported by DDVNs using quorums. The proposed approach allows individual nodes that are corrupted due to an attack to be detected or excluded when the DNN produces an output. Metrics such as corruption factor and success probability of an attack are considered for evaluating the security aspects of DNNs. A simulation study demonstrates that if the number of corrupted nodes is less than a given threshold for decision-making in a quorum, the \texttt{QUDOS} approach always prevents attacks. Furthermore, the study shows that increasing the size of the quorum has a better impact on security than increasing the number of layers. One merit of \texttt{QUDOS} is that it enhances the security of DNNs without requiring any modifications to the algorithm and can therefore be applied to other classes of problems.  
\end{abstract}


\maketitle

\section{Introduction}
\label{sec:introduction}

Industry 4.0 is a digital transformation~\cite{ustundag2017industry} of the industry by connecting sensors, machines and production systems to communicate with each other. Data from sensors, machines and processes are exchanged between different entities and are linked using rule-based approaches or machine learning (ML) based algorithms, such as Deep Neural Networks (DNNs) to optimize processes and predictions. The quality of the data is crucial for training DNNs to achieve high-quality models for inference tasks~\cite{bosu2013data}. 

A wide variety of use cases within the industry are underpinned by DNN-based inference. Examples include (a) visual inspection in manufacturing, (b) navigation of robots and path planning for smart machines, and (c) conversational user interfaces~\cite{7841045, schonle2021industry}. The quality of data for inference is guaranteed by a quality analysis during data pre-processing. 

In this paper, the focus is on \texttt{QUDOS}, a Distributed Data Validation Network (DDVN) extended by quorums that implements a DNN. The DDVN solves the problem of a single point of failure in data pre-processing systems by distributing the pre-processing tasks onto different nodes located on resources in a shop floor, edge or cloud data center. 

In this context, the \ac{DDV} is used to validate the quality of data gathered, for example, from a shop floor in a smart manufacturing setting. The quality thresholds are improved by including a quality score with the data required by all entities. The data score determines whether additional data or more accurate data is required for prediction. A problem occurs when a DDVN implements a sequential algorithm, like the DNN, where individual layers, each corresponding to a node, are corrupted. Due to a corrupted layer, the computations of all layers after it will be affected. To prevent the influence of individual corrupted layers, the DDVN is extended by using a quorum-based approach. In the quorum approach, each layer in the DNN is replicated on other nodes and also used simultaneously for performing computations. The set of nodes that calculate the same layer is called the quorum. In order to make a decision in a quorum, a defined set of nodes (threshold) must publish the same result. Such a quorum-based approach ensures that an attacker must corrupt several nodes to be successful in altering the output produced by the DNN.

This paper makes the following research contributions:

(i) This research will for the first time map the layers of a \ac{DNN} to validators in a \ac{DDV} for enhancing the resilience of data driven processes running across resources on a manufacturing shop floor, the edge and the cloud.

(ii) A novel quorum-based approach is developed and implemented for mapping a \ac{DNN} on to a \ac{DDV} so that the security of the overall system is enhanced in the face of failures or attacks on validators in a \ac{DDV}.

(iii) A theoretical and simulation-based evaluation of the attack success probability of a \ac{DNN} and an analysis of the attack success probability demonstrating the benefits of the quorum-based enhancement. 

The simulation study was carried out by taking into account different quorum sizes (number of nodes in a quorum), layer counts, quorum counts (number of layers secured by a quorum), corrupted node counts and threshold sizes. The key observations are (a) a single corrupted node has no impact when all layers are secured by a quorum, (b) increasing the quorum size is better than increasing the layer size, and (c) the threshold has a high impact on the success probability of an attack. 

The remainder of this paper is organized as follows. 
Section~\ref{sec:relatedwork} discusses the related work. 
Section~\ref{sec:qudos} presents \texttt{QUDOS} and the underlying approach for implementing a \ac{DNN} with a \ac{DDV} and the quorum-based extension. 
Section~\ref{sec:analysis} considers an experimental analysis of the corruption factor of a \ac{DNN} and the success probability of an attack.
Section~\ref{sec:outlook} concludes this paper by highlighting open research questions.

\section{Related Work}
\label{sec:relatedwork}

The following section describes research related to approaches for distributing \ac{DNN} and enhancing the security of distributed machine learning.  

\subsubsection*{Distributing Deep Neural Networks}
Deep Neural Networks (DNNs) are often employed in computer vision applications (for example, in automated smart factory floors for enhancing production efficiency and quality)~\cite{dnn-01, dnn-02}. 
DNNs are composed of a sequence of layers, such as convolution, activation or pooling. 
Recent research has demonstrated that DNN layers can be partitioned and distributed across different compute resources for efficiency~\cite{lockhart2020scission, neurosurgeon, edge-dnn-survey-1}.

The distribution of the layers of a DNN across multiple resources has the potential to enhance the resilience of the overall system. Thus single point failures can be reduced as the entire DNN will not fail if one compute resource fails or malfunctions due to an attack. 
High-value and business-critical operations can be rapidly recovered on a factory floor with minimum disruption if the layers of a DNN are replicated (for redundancy) within a network of computing resources. 

There are five categories of techniques that are used to distribute DNN based on its layers. They are: (i)~Profiling and estimation based~\cite{neurosurgeon, deepwear}, (ii)~linear programming based~\cite{jalad, jointdnn}, (iii)~DNN structural modification based~\cite{deepthings, modnn}, (iv)~Approximation based~\cite{dynamicdnnsurgery}, and (v)~Benchmarking based~\cite{lockhart2020scission, lavea}.

The research in this paper builds on benchmarking-based DNN partitioning and distribution, namely Scission~\cite{lockhart2020scission, ahn2021scissionlite}, to map DNN layers onto compute resources in a network. 
Benchmarking-based techniques are based on actual measurements and no assumptions are made regarding the underlying hardware or  performance of the layers on the compute resource. Operational parameters (such as load on the network and computing resource) are taken and the optimal partition is calculated. This approach requires no modification to the DNN code and has been demonstrated to work on production DNNs with many layers (and layer types and configurations).

\subsubsection*{Security of Distributed Machine Learning}
Within machine learning research, the focus is usually on optimizing the prediction accuracy and the learning process. Security by design, specifically making machine learning resilient to external attacks, is minimally considered or taken for granted. 
Security enhancement mechanisms such as the use of a Public Key Infrastructure (PKI) for issuing certificates, with which authenticity can be guaranteed, or the encryption of messages using Transport Layer Security (TLS) can be applied to distributed machine learning. However, the applicability of these mechanisms within specific domains or their advantages and disadvantages are not fully known.

A mechanism based on blockchain and diversity to improve the robustness of machine learning was introduced~\cite{Shukla2021}. Different results for a problem are calculated by varying data and machine learning algorithms. Subsequently, the different prediction results are aggregated using a consensus algorithm. The research presented in this paper does not employ blockchains and thus saves overhead time. Instead, the research in this paper supports rule-based approaches and protects the DNN using a quorum approach. Furthermore, the data quality by enhancing the data-preprocessing step is improved. 

In addition to the known protection mechanisms used in distributed systems, some approaches make the machine learning algorithm itself more robust against attacks. These approaches are algorithm-specific and are not generic. %
Therefore, they cannot be easily applied to other machine learning algorithms.

For example, a gradient descent machine learning algorithm that handles byzantine failures is presented~\cite{BlanchardMGS17, 10.1145/3154503}. A practical application is Google\textquotesingle s Federated Learning where $m$ worker machines analyze $\frac{N}{m}$ data samples, where $N$ is the total number of samples. The approach is limited to gradient descent learning algorithms. 

The Nash Equilibrium is used to improve the security and resilience of distributed support vector machines~\cite{7266621, Zhang2021}. A two-player machine learning algorithm is assumed in which the first player is the attacker who attempts to break the learning process and the second player is the learner who aims to optimize the trained model so that the prediction results are optimal. The resulting min-max problem is solved using ADMoM (Alternating Direction Method of Multipliers) technique. Furthermore, the importance of the network topology is shown by comparing line and star topology. The more neighbors a node has the more sensitive it is to an attack. This approach is also limited to the distributed support vector machines.

The security of DNNs mostly concerns adversarial attacks~\cite{jakubovitz2018improving,liu2019feature}. In these attacks, it is assumed that the attacker manipulates the input data of the DNN and thus falsifies the prediction. The manipulated input data is usually not recognizable to a human, which is why it is difficult to detect. By using a denoising ensemble and verification ensemble~\cite{9006090}, any existing noise can be filtered and the attack prevented. However, these attacks always assume that the attacker is outside the DNN. In contrast, this paper considers the impact of corrupted layers inside the DNN and does not take attacks from outside into account.

\subsubsection*{Security by Distributed Data Validation Network}
The Distributed Data Validation Network (DDVN)~\cite{8904039,9210350,wallis2020corrupted} is used in environments where data quality should be improved utilizing data preprocessing. The focus is on the security goals of integrity and availability. During data preprocessing, different machine learning algorithms are used to determine the quality of the data. It is essential that pre-processing is not carried out on a single node within the system. Otherwise, a successful attack can manipulate the entire data in the pre-processing phase. Here the DDVN can be employed; it addresses the problem of a single point of failure by distributing the pre-processing phase onto different nodes.

In this paper, a DNN is implemented and analyzed using \texttt{QUDOS}, a DDVN extended with quorums. The \texttt{QUDOS} approach differs from any of the above-mentioned approaches in the following ways: (i) it mitigates the problem of a single point of failure by adopting a completely distributed approach, (ii) it employs existing resources available in the computing continuum (cloud, edge, and shop floor), which may not be necessarily trusted, and (iii) it can be used as a generic approach for other sequential machine learning algorithms with minimum modification.

\begin{figure*}[h]
	\centering
    \includegraphics[width=\textwidth]{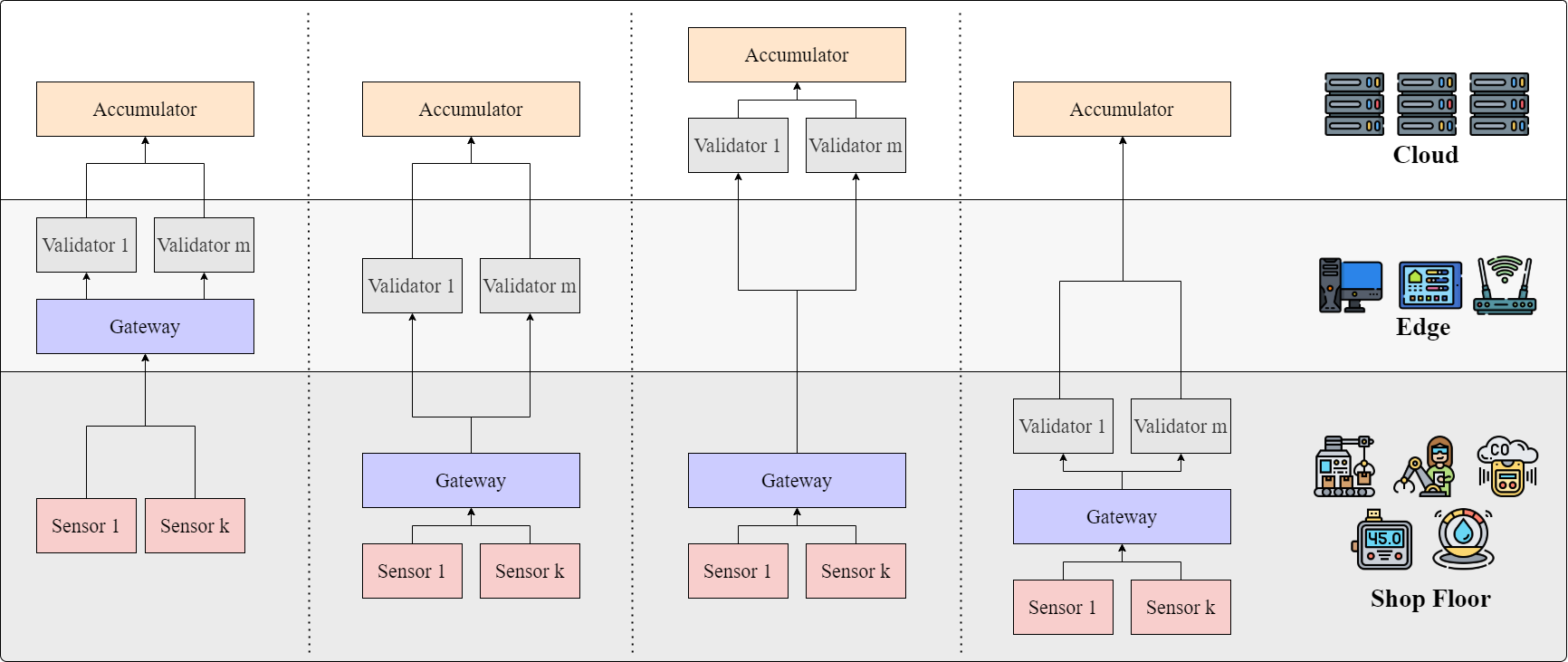}
    \caption{Different implementations of the DDVN, distributed across the three layers (shop floor, edge and cloud) in an Industry 4.0 environment}
    \label{fig:usecase}
\end{figure*}

\section{QUDOS}
\label{sec:qudos}

In this section, the \ac{DDV}, its structure and positioning within an industrial environment are considered. The \ac{DNN} and a practical implementation by using a \ac{DDV} are presented. Furthermore, the quorum-based approach to implement a robust and secure \ac{DNN}, referred to as \texttt{QUDOS}, is considered.

\subsubsection*{Background}
Data pre-processing is used to prepare and evaluate data for data analysis. Missing attributes and missing values, improper data types, out-of-range data, unnecessary data, incomplete data and corrupted data~\cite{famili1997data} are some of the problems that are considered by an ideal data pre-processing method. 

The \ac{DDV}~\cite{8904039} solves the single point of failure that is inherent to a standalone data pre-processing node. It distributes the data validation task over multiple validation nodes so that the impact of a single corrupted node is minimal. Figure~\ref{fig:usecase} shows the structure of the \ac{DDV} and four different implementations by distributing it over the cloud, edge and shop floor. Multiple sensors of a machine (for example, temperature sensors and humidity sensors) forward their data to gateways. The gateways distribute the received sensor data to the correct validators and the associated accumulators. For this, the gateways receive a list of the correct validators $\{v_1, v_2, .., v_m\}$ from the accumulators. Each validator carries out the validation task and sends the validation result to the appropriate accumulator. The accumulator aggregates the validation results and makes decisions about the correctness and trustworthiness of the received data. 

To enhance \ac{DDV} clusters are considered. A cluster is a logical grouping of validators based on similar characteristics. These characteristics include: (a) software properties, such as the operating system used, (b) hardware properties, for example, whether a GPU is available, and (c) validation capabilities, for example, can assess the temperature of a machine.

In an industrial environment (see Figure~\ref{fig:usecase}), the sensors are located on the machine or other devices on the shop floor. The gateway is either the sensor itself, provided it is network-enabled, has sufficient computing power and is trusted, or a specific hardware component located near the sensor. The validators may be any available hardware on the shop floor, for example, a terminal located near the machine or an operator's tablet on which production orders are displayed. In addition to resources available on the shop floor, a validator may also run on edge resources (for example, routers or micro clouds) and cloud data center resources. Since the clusters are only logical boundaries additional hardware is not required. Typically, the accumulators are located in a protected area, for example in the company's internal cloud where sensitive production, customer and company data is processed and stored. 

Communication between the components of a \ac{DDV} is essential. Unidirectional communication between validators and accumulators is considered. The different possibilities of communication are: 

\begin{itemize}
    \item One validator is connected to one accumulator and the validator forwards its result to the accumulator.
    \item A set of validators are connected in series and the last validator in the series forwards the result to the accumulator.
    \item A set of validators are connected in parallel and each validator forwards its result to the accumulator.
    \item A set of validators are connected in serial and parallel and end with a serial validator so that only one validator forwards the result to the accumulator.
    \item A set of validators are connected in serial and parallel and end with parallel validators so that multiple validators forward their results to the accumulator.
\end{itemize}

Independent of the type of communication, the integrity of the intermediate results of the validators must be secured. For this purpose, each validator has its keypair and uses it to sign the result. The succeeding validators can thus not corrupt the intermediate results. One possible framework for implementing non-repudiation and integrity is the blockchain (outside the scope of this paper). 

The chain of trust can either be verified by (a) the successive validator and thus be terminated at an early stage in case of a verification failure or (b) only at the end of a validation chain by the accumulator. Using (a) has the disadvantage that a corrupted validator can drop results from previous validators. 

\subsubsection*{Motivation}
The individual layers of a sequential DNN depend on the preceding layers, Therefore, if the data of a layer is corrupted due to an attack then it has an impact on all subsequent layers. Thus, the final result may also be negatively impacted when the data in any layer of the DNN is corrupted. As a first step, using the DDVN, the dependencies that exist among the validation nodes are decoupled, so that a set of nodes can calculate a certain layer. This ensures that a successful attack on one node does result in a single point of failure. 

\begin{figure}[h]
    \centering
    
    \begin{subfigure}[h]{\linewidth}
	\centering
    \begin{tikzpicture}[every node/.style={fill=white, draw=black}, align=center, scale=0.9, transform shape]
    

        \def\ylayer{0}
        
        
         \node (gateway_2) [rectangle_blue_vertical, inner xsep=20.5] at (0.0, \ylayer) {Gateway};
        
        \node (layer_1) [rectangle_gray_vertical, inner xsep=20.5] at (2.0, \ylayer) {Layer $1$};
        \node (layer_2) [rectangle_gray_vertical, inner xsep=20.5] at (4.0, \ylayer) {Layer $2$};
        \node (layer_i) [rectangle_none] at (4.5, \ylayer + 1.5) {...};
        \node (layer_n) [rectangle_gray_vertical, inner xsep=20.5] at (6.0, \ylayer) {Layer $m$};

        \node (accumulator_2) [rectangle_orange_vertical] at (8.0, \ylayer) {Accumulator};

        
        \draw[->, dashed] (gateway_2.south) -- (layer_1.north);
        \draw[->, dashed] (layer_1.south) -- (layer_2.north);
        \draw[->, dashed] (layer_n.south) -- (accumulator_2.north);

	\end{tikzpicture}
	\caption{\ac{DNN} with sequential structure}
	\label{fig:dnn_sequential}
	\end{subfigure}
	
    \begin{subfigure}[h]{\linewidth}
	\centering
    \begin{tikzpicture}[every node/.style={fill=white, draw=black}, align=center, scale=0.9, transform shape]
    
    
        \def\ypasses{0}
        
        
        \node (gateway_1) [rectangle_blue_vertical, inner xsep=20.5] at (0.0, \ypasses) {Gateway};
        
        \node (cluster) [cloud, draw, cloud puffs=10, cloud puff arc=120, cloud ignores aspect, minimum width=6.5cm, minimum height=3.5cm] at (3.5,\ypasses + 1.5) {};
        
        \node (validator_1) [circle_gray] at (1.5, \ypasses + 1.0) {$v_1$};
        \node (validator_2) [circle_gray] at (3.5, \ypasses + 2.3) {$v_2$};
        \node (validator_m) [circle_gray] at (5.5, \ypasses + 1.0) {$v_m$};
        \node (validator_4) [circle_gray, opacity=0.3] at (5.1, \ypasses + 2.3) {$v_3$};
        \node (validator_5) [circle_gray, opacity=0.3] at (3.7, \ypasses + 1.0) {$v_4$};
        \node (validator_6) [circle_gray, opacity=0.3] at (2.0, \ypasses + 2.2) {$v_5$};
        
        \node (accumulator_1) [rectangle_orange_vertical] at (8.0, \ypasses) {Accumulator};


        \draw[->, dashed] (gateway_1.south) -- (validator_1.west);
        \draw[->, dashed] (validator_1.east) -- (validator_2.west);
        \draw[->, dotted] (validator_2.east) -- (validator_m.west);
        \draw[->, dashed] (validator_m.east) -- (accumulator_1.north);

	\end{tikzpicture}
	\caption{Implementation of a sequential \ac{DNN} with a \ac{DDV}}
	\label{fig:ddvn_sequential}
	\end{subfigure}
	
	\begin{subfigure}[h]{\linewidth}
	\centering
    \begin{tikzpicture}[every node/.style={fill=white, draw=black}, align=center, scale=0.9, transform shape]
    
    
        \def\ypasses{0}
        
        
        \node (gateway_1) [rectangle_blue_vertical, inner xsep=20.5] at (0.0, \ypasses) {Gateway};
        
        \node (cluster) [cloud, draw, cloud puffs=10, cloud puff arc=120, cloud ignores aspect, minimum width=6.5cm, minimum height=4.5cm] at (3.5,\ypasses + 1.5) {};
        
        \node (validator_1a) [circle_gray_min] at (1.6, \ypasses + 2.5) {$v_1'$};
        \node (validator_1b) [circle_gray_min] at (1.6, \ypasses + 1.5) {$v_1''$};
        \node (validator_1c) [circle_gray_min] at (1.6, \ypasses + 0.5) {$v_1'''$};
        
        \node (validator_2a) [circle_gray_min] at (3.7, \ypasses + 2.1) {$v_2'$};
        \node (validator_2b) [circle_gray_min] at (3.7, \ypasses + 1.2) {$v_2''$};
        \node (validator_2c) [circle_gray_min] at (3.7, \ypasses + 0.2) {$v_2'''$};
        
        \node (validator_m) [circle_gray_min] at (5.5, \ypasses + 1.0) {$v_m$};
        \node (validator_4) [circle_gray_min, opacity=0.3] at (5.1, \ypasses + 2.3) {$v_3$};
        \node (validator_5) [circle_gray_min, opacity=0.3] at (4.0, \ypasses + 3.0) {$v_4$};
        \node (validator_6) [circle_gray_min, opacity=0.3] at (2.5, \ypasses + 2.8) {$v_5$};
        
        \node (accumulator_1) [rectangle_orange_vertical] at (8.0, \ypasses) {Accumulator};


        \draw[->, dashed] (gateway_1.south) -- (validator_1a.west);
        \draw[->, dashed] (gateway_1.south) -- (validator_1b.west);
        \draw[->, dashed] (gateway_1.south) -- (validator_1c.west);
        
        \draw[->, dashed] (validator_1a.east) -- (validator_2a.west);
        \draw[->, dashed] (validator_1a.east) -- (validator_2b.west);
        \draw[->, dashed] (validator_1a.east) -- (validator_2c.west);
        
        \draw[->, dashed] (validator_1b.east) -- (validator_2a.west);
        \draw[->, dashed] (validator_1b.east) -- (validator_2b.west);
        \draw[->, dashed] (validator_1b.east) -- (validator_2c.west);
        
        \draw[->, dashed] (validator_1c.east) -- (validator_2a.west);
        \draw[->, dashed] (validator_1c.east) -- (validator_2b.west);
        \draw[->, dashed] (validator_1c.east) -- (validator_2c.west);
        
        \draw[->, dotted] (validator_2a.east) -- (validator_m.west);
        \draw[->, dotted] (validator_2b.east) -- (validator_m.west);
        \draw[->, dotted] (validator_2c.east) -- (validator_m.west);
        
        \draw[->, dashed] (validator_m.east) -- (accumulator_1.north);

	\end{tikzpicture}
	\caption{Implementation of a sequential \ac{DNN} with \texttt{QUDOS}}
	\label{fig:qudos_sequential}
	\end{subfigure}
	
    \caption{Implementation of a sequential \ac{DNN} by using a \ac{DDV} and \texttt{QUDOS}}
    \label{fig:dnn_ddvn_sequential}
\end{figure}
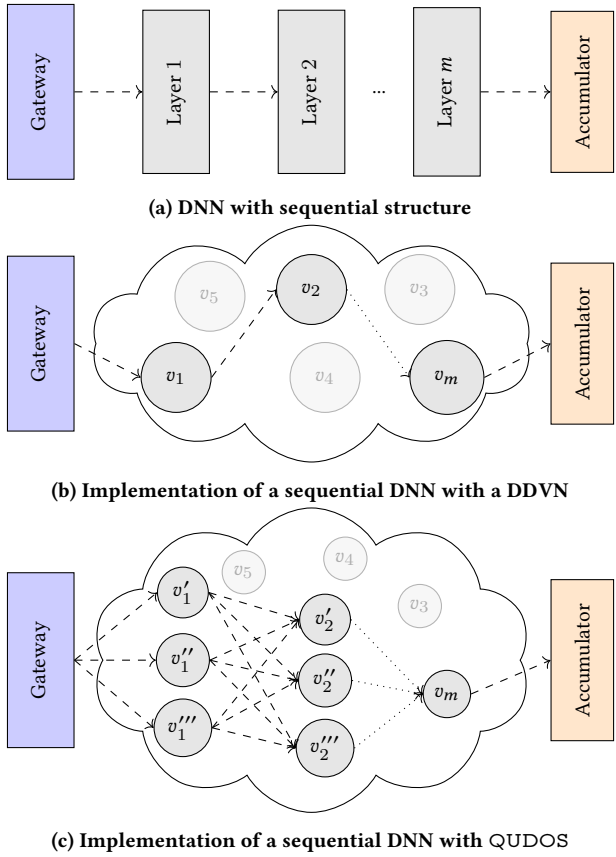

A DNN is a sequence of layers such that given an image or video input, each layer performs a sequence of different operations. The most common layers in a DNN are fully connected, convolution, pooling, activation and softmax. There are two categories of DNNs, namely sequential and branched DNNs. In a sequential DNN (see Figure~\ref{fig:dnn_ddvn_sequential}), the input of one layer is connected to the next, resulting in a series of layers forming the neural network. The mapping of a DNN to DDVN is shown in Figure~\ref{fig:ddvn_sequential}. 
In a branched DNN, however, multiple concurrent branches of layers will be connected. In other words, a layer may be connected to more than two layers resulting in parallel paths. Using \texttt{QUDOS}, both sequential and branched \ac{DNN} can be implemented. In a sequential \ac{DNN} (see Figure~\ref{fig:qudos_sequential}), one validator is connected to a single subsequent validator at a time. In comparison, with a branched \ac{DNN}, the result of a single validator is simultaneously forwarded to two subsequent validators.

\subsubsection*{QUDOS: Quorum-Based DDVN}
If an attacker is successful in attacking a node, then the resulting corruption in data will adversely impact the final result obtained from the DNN. \texttt{QUDOS} proposed in this paper mitigates this challenge as it extends the existing DDVN with quorums of certain layers. Quorums are used in fault-tolerant and highly available distributed systems to ensure consistency~\cite{vukolic2013origin}. A quorum is defined as the minimum number of votes $n_{min} \in \mathbb{N}$ that are necessary to perform an operation and $n_{min} \leq n$, where $n$ are all votes. In terms of \texttt{QUDOS}, a quorum approach means replicating a layer on $n$ nodes, where $n_{min}$ nodes must have the same result for a quorum decision: 

\begin{equation}
    n_{min} > t \cdot n
\end{equation}

Where $\Set{t | 0 < t \leq 1}$ is the threshold. This allows a validator to be replaced by a quorum and an attack must corrupt $t \geq \frac{1}{2}$ of all participating validators in a quorum. However, even if an attack reaches the majority in a quorum, as long as the attack does not corrupt $100\%$ in the quorum, an anomaly can be detected. By increasing the threshold $t$, the success of an attack can be decreased in a quorum-based \ac{DDV} (see Section~\ref{sec:analysis}). In a high-security environment, $t$ is set $1$, so a single deviation between the quorum participants is recognized.

\subsubsection*{Chain of Trust in QUDOS}
\begin{figure}[h]
    \centering
    
    \begin{subfigure}[h]{\linewidth}
	\centering
    \includegraphics[width=\textwidth]{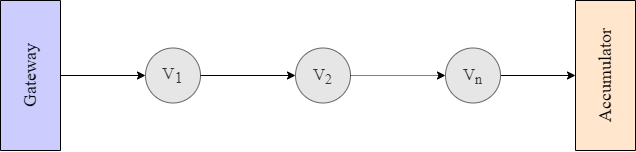}
	\caption{Sequential chain of trust}
	\label{fig:communication_1}
	\end{subfigure}
	
    \begin{subfigure}[h]{\linewidth}
	\centering
    \includegraphics[width=\textwidth]{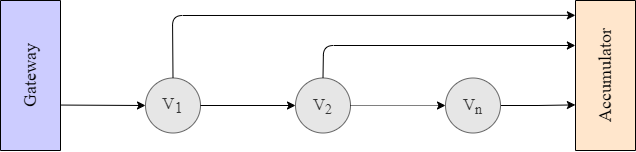}
	\caption{Trust by the accumulator}
	\label{fig:communication_2}
	\end{subfigure}
	
    \begin{subfigure}[h]{\linewidth}
	\centering
    \includegraphics[width=\textwidth]{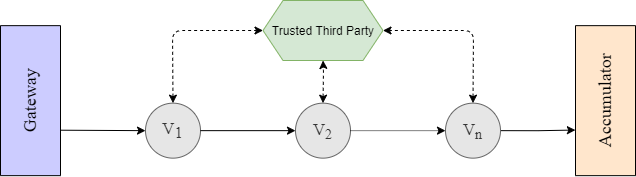}
	\caption{Trust by a Trusted Third Party}
	\label{fig:communication_3}
	\end{subfigure}
	
    \caption{Comparison of different chain of trust approaches} 
    \label{fig:chain_of_trust}
\end{figure}

Trust is a basic prerequisite for the successful cooperation in distributed systems. Otherwise, data and results from other nodes can only be used with limitations for processing tasks and accumulating overall results. The same applies to the use of a DNN or \texttt{QUDOS}, if the results from a previous layer cannot be trusted, then the succeeding layer cannot calculate a reliable result. This behavior can be compared with a chain of trust in certificate chains. In certificate chains as well as sequential DNNs, trust must be ensured from the root certificate, also called root certificate (first layer), to the last entity certificate (last layer). Figure~\ref{fig:chain_of_trust} shows a comparison of three different approaches, namely Case (a), Case (b) and Case (c), for ensuring the chain of trust from previous layer results.

\begin{itemize}
    \item In Case (a), each validator $v$ signs its result and the hash value of the result received from the previous validator, as long as the result and the given signature correspond to its expectation. Otherwise, error handling must be performed, which can be e.g. notification of the service personnel via email. If the trust chain was successfully established, it will be ensured that the expected layers of the DNN were used for calculating the prediction result.
    \item Case (b) shows an approach in which the intermediate results are forwarded to the subsequent validator $v_{i+1}$ as well as directly to the accumulator. The accumulator can thus check the intermediate results, and in the case of an error, compare them with the input data from the subsequent validator. Furthermore, no chain of trust has to be established between the validators, as this is done directly via the accumulator. 
    \item The third approach, Case (c), describes the use of a Trusted Third Party~\cite{abadi2004trusted}. The central point here is that the accumulator trusts the Trusted Third Party. Thus, the trust relationship can be outsourced from the Accumulator to the Trusted Third Party. The Trusted Third Party signs each layer result and is responsible for preventing untrusted nodes from taking over the identity of a participating node from the sequential process. It should be noted that the Trusted Third Party must be distributed to different nodes, otherwise, there will be another single attack surface in the DDVN. 
\end{itemize}

The three trust approaches presented do not influence the correctness of the results of a DNN. Furthermore, the approaches only ensure that no unknown participants or even known nodes add or remove additional results to the DNN.

\subsubsection*{Metrics}
In the following, two security metrics (corruption factor and attack success probability) for a DNN are defined and the equation for calculating the selection probability of corrupted nodes inside a DDVN is given. These metrics are used for the security analysis of \texttt{QUDOS} in Section~\ref{sec:analysis}.

The \textit{corruption factor} defines how much a system (DNN, Decision Tree, etc.) is corrupted if a sub-component has been corrupted. In the given context, a layer of a DNN is the sub-component to be considered. The corruption factor of a Decision Tree~\cite{wallis2020corrupted}, is used as the basis for defining the corruption factor of a DNN. Therefore, the corruption factor $\eta$ of a single layer $l$ is the ratio of passes $n_l$ to all passes $n$. In a sequential \ac{DNN} for every decision process, each layer has to be passed, which leads to a corruption factor of $1$ (see Figure~\ref{fig:dnn_sequential}).

\begin{equation}
\label{eq:corruption_factor}
   \eta(l) = \frac{n_l}{n} = 1
\end{equation}

\begin{figure}[h]
    \centering
    
    \begin{subfigure}[h]{\linewidth}
	\centering
    \includegraphics[width=\textwidth]{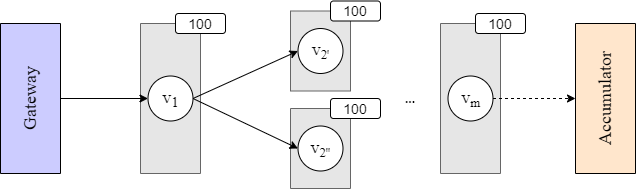}
	\caption{Branched DNN structure and number of visits per layer and node}
	\label{fig:dnn_branched}
	\end{subfigure}
	
    \begin{subfigure}[h]{\linewidth}
	\centering
    \includegraphics[width=\textwidth]{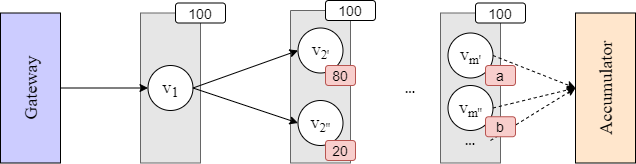}
	\caption{Decision Tree structure and number of visits per layer and node}
	\label{fig:decision_tree}
	\end{subfigure}

    \caption{Comparison of Branched DNN and Decision Tree structure and number of visits per layer and node} 
    \label{fig:metrics}
\end{figure}

The given corruption factor $\eta$ must be adapted for layers that consist of several sub-layers since a branched \ac{DNN} behaves differently than a Decision Tree. In a Decision Tree structure, the number of visits of a layer is constant and also the same across all layers even when there are multiple nodes inside a layer. In Figure~\ref{fig:metrics} the comparison of the Branched DNN to the Decision Tree is given. The grey boxes correspond to one layer each and the visits ($100$) per layer are shown in white. In the Branched DNN, each branched layer is visited during a prediction process. Thus, in Case (a), both nodes $v_{2'}$ and $v_{2''}$ have the same number of visits ($100$). In Case (b), a Decision Tree, each layer is visited the same number of times as the other layers, but only one node in a layer is used for a single prediction process. So, in the given example, $v_{2'}$ is visited $80$ times out of $100$ predictions and $v_{2''}$ is visited the remaining $20$ times.   

For a \ac{DNN} with a branched layer, the number of visits increases proportionally with the number of sub-layers $s$. Therefore, the corruption factor for a sub-layer is defined as:

\begin{equation}
\label{eq:corruption_factor_sublayer}
   \eta(l_{sub}) = \frac{1}{s}
\end{equation}

When a set of layers $L$ is chosen from the \ac{DNN} then the corruption factor is defined as:

\begin{equation}
\label{eq:corruption_factor_validators}
   \eta(L) =  1 - \prod_{l \in L} 1-\eta(l)
\end{equation}

The corruption factor $\eta$ of the \ac{DNN} corresponds to the success probability $\gamma$ of an attack. Therefore, the \textit{attack success probability} is defined as follows:

\begin{equation}
\label{eq:success_probability_validators}
    \gamma(V) = \eta(V)
\end{equation}

The set of all existing validators $V$ has cardinality $n=|V|$. Among all validators, there is a set of corrupted validators $C \subseteq V$ with $c=|C|$. The total number of validators chosen for the DNN is $m$. Thus, the probability that there are $x$ corrupted validators in the chosen validators is: 

\begin{equation}
    p = \frac{\binom{c}{x} \cdot \binom{n-c}{m-x}}{\binom{n}{m}}
\end{equation}

\section{Analysis}
\label{sec:analysis}
In this section, the security of distributed DNNs using the \texttt{QUDOS} approach is considered. First, it is shown what knowledge an attacker can use to attack a DNN and what the corresponding impacts are. Subsequently, the advantage of using multiple nodes (DDVN approach) is discussed and it is shown that by increasing the number of nodes, the probability of success for the attacker decreases, since the selection of corrupted nodes for the machine learning algorithm becomes less. Finally, the \texttt{QUDOS} approach is analyzed under the assumption that all nodes are used for the DNN. Among other things, it is shown that as long as an attacker has only corrupted one node, the probability of success is zero.

\subsubsection*{Attacker Knowledge}
An attack on a system involves a wide variety of factors that give an attacker an advantage and must therefore be considered in terms of security. Factors that must be considered before an attack occurs are a) the capabilities of an attacker, b) the hardware available for an attack and c) the knowledge of an attacker. Especially when considering the security of the \texttt{QUDOS} approach, the knowledge of an attacker has a significant impact. The type of knowledge an attacker may have are:

\begin{itemize}
    \item \textbf{Oblivious:} The attacker has no knowledge about the layer, structure, expected result or data inside the DDVN. Therefore, a successful attack can return any result and the impact of the final prediction result is unknown.
    \item \textbf{Expected Result of the Layer is known:} The attacker knows the expected result, for example, by using the corrupted layer's usual operations. So, the attacker can invert or forward the expected result but does not know the impact on the overall result as long as the corrupted layer is not the final layer of the \ac{DNN}.
    \item \textbf{Expected Result of the DNN is known:} The attacker knows the expected final result and can try to reverse engineer the subsequent layers. In the worst-case scenario for the DNN, the inter-layer results were successfully reverse-engineered.
    \item \textbf{Input Data of the DNN is known:} The attacker knows the input data for the DNN but as long as the attacker does not have any additional information the impact of the final prediction result is unknown.
    \item \textbf{DNN Structure is known:} An attacker may be able to determine appropriate nodes that the layers of a sequential DNN are mapped onto. The final layer of the \ac{DNN} is usually the best target for an attacker because this layer returns the final prediction result. If the corrupted layer is an intermediate layer the attacker will be able to control the whole \ac{DNN} when the weights and complete structure of the succeeding layers are known. For algorithms that follow varying sequential flows, e.g. a decision tree, the attacker also needs to know which path is followed in the decision-making process. 
\end{itemize}

This paper assumes that the attacker has no knowledge (oblivious) in the subsequent analysis.

\subsubsection*{Enhancing Security with additional Nodes}
When a sequential \ac{DNN} is implemented with a \ac{DDV} each validator corresponds to a layer. The layer for the input data is the gateway and the final layer is placed on a validator. This allows the validation result to be distributed to several different accumulators.

\begin{figure}[tbh]
	\centering
\begin{tikzpicture}
\begin{axis}[
height=6cm,
width=8.5cm,
grid=both, minor tick num=1,
legend columns=1,
legend style={at={(0.85,0.70)},
	anchor=north},
ylabel={Selection Probability $p$},
xlabel={Layer Count $l$},
      ymin=0,ymax=0.41,
      xmin=50,xmax=1000
]

\addplot[dashed, mark=triangle*,color1a, mark options={solid}] coordinates {
( 50 , 0.0004999999999999449 )
( 150 , 0.0014999999999999458 )
( 250 , 0.0024999999999999467 )
( 500 , 0.0050000000000000044 )
( 750 , 0.007499999999999951 )
( 1000 , 0.010000000000000009 )
};
\addlegendentry{$c=1$}

\addplot[dashed, mark=diamond*,color2a, mark options={solid}] coordinates {
( 50 , 0.0024975511512587145 )
( 150 , 0.007477682831183441 )
( 250 , 0.012437903571548392 )
( 500 , 0.02475173696749189 )
( 750 , 0.036942430726920494 )
( 1000 , 0.049010910724489265 )
};
\addlegendentry{$c=5$}

\addplot[dashed, mark=x,color3a, mark options={solid}] coordinates {
( 50 , 0.004988988990570875 )
( 150 , 0.014899819922756241 )
( 250 , 0.024721716829721907 )
( 500 , 0.048892020419913584 )
( 750 , 0.07252187058849946 )
( 1000 , 0.09562203607115127 )
};
\addlegendentry{$c=10$}

\addplot[dashed, mark=o,color4a, mark options={solid}] coordinates {
( 50 , 0.02470216435165884 )
( 150 , 0.07232582743377058 )
( 250 , 0.11766830549803187 )
( 500 , 0.22173536851220166 )
( 750 , 0.3137449456303263 )
( 1000 , 0.39506881492826973 )
};
\addlegendentry{$c=50$}

\end{axis}
\end{tikzpicture}
	\caption{Probability $p$ of selecting one or more of the corrupted nodes $c$ for the DNN with layer count $l=[50, 150, 250, 500, 750, 1000]$ and the number of nodes of the DDVN $n=100,000$}
	\label{fig:analysis_0b}
\end{figure}
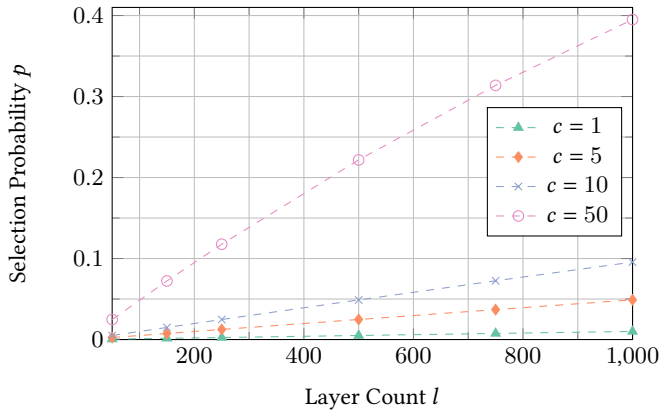

Figure~\ref{fig:analysis_0a} and Figure~\ref{fig:analysis_0b} show the values for the selection probability of corrupted nodes. Two different numbers for validators inside the DDVN ($100$ and $100,000$) are assumed and the number of corrupted validators $c$ and the layers $l$ are varied. By considering sequential \ac{DNN}, it follows that when one or more corrupted validators are selected, the corruption factor $\eta = 1$. As the number of layers increases, the selection probability of a corrupted node also increases. Therefore, in a sequential \ac{DNN}, the number of layers should be kept to a minimum, because $\lim_{l \to n} \eta(l) = 1$. Furthermore, despite the low ratio of corrupted validators to non-corrupted validators, the selection probability for a corrupted node is still very high and consequently also the probability of corrupting the complete DNN. For the given figures, it was assumed that each node can calculate each layer of the \ac{DNN}. 

\begin{figure}[tbh]
	\centering
\begin{tikzpicture}
\begin{axis}[
height=6cm,
width=8.5cm,
grid=both, minor tick num=1,
legend columns=2,
legend style={at={(0.75,0.90)},
	anchor=north},
ylabel={Selection Probability $p$},
xlabel={Layers Count $l$},
      ymin=0,ymax=1.05,
      xmin=1,xmax=10
]

\addplot[dashed, mark=triangle*,color1a, mark options={solid}] coordinates {
( 1 , 0.010000000000000009 )
( 2 , 0.020000000000000018 )
( 3 , 0.030000000000000027 )
( 4 , 0.040000000000000036 )
( 5 , 0.050000000000000044 )
( 6 , 0.06000000000000005 )
( 7 , 0.06999999999999995 )
( 8 , 0.07999999999999996 )
( 9 , 0.08999999999999997 )
( 10 , 0.09999999999999998 )
};
\addlegendentry{$c=1$}

\addplot[dashed, mark=diamond*,color2a, mark options={solid}] coordinates {
( 1 , 0.050000000000000044 )
( 2 , 0.09797979797979794 )
( 3 , 0.14400123685837973 )
( 4 , 0.18812488444299935 )
( 5 , 0.23041004671159315 )
( 6 , 0.2709147810951935 )
( 7 , 0.3096959097603428 )
( 8 , 0.3468090328915071 )
( 9 , 0.3823085419734904 )
( 10 , 0.41624763307384804 )
};
\addlegendentry{$c=5$}

\addplot[dashed, mark=x,color3a, mark options={solid}] coordinates {
( 1 , 0.09999999999999998 )
( 2 , 0.19090909090909092 )
( 3 , 0.273469387755102 )
( 4 , 0.34836945087313276 )
( 5 , 0.41624763307384804 )
( 6 , 0.47769525064502194 )
( 7 , 0.5332595856827855 )
( 8 , 0.5834467270072172 )
( 9 , 0.6287242566803458 )
( 10 , 0.6695237889132748 )
};
\addlegendentry{$c=10$}

\addplot[dashed, mark=o,color4a, mark options={solid}] coordinates {
( 1 , 0.5 )
( 2 , 0.7525252525252525 )
( 3 , 0.8787878787878788 )
( 4 , 0.9412683536394877 )
( 5 , 0.9718577527855878 )
( 6 , 0.9866694618458047 )
( 7 , 0.9937601736299512 )
( 8 , 0.9971149189901924 )
( 9 , 0.9986828977998705 )
( 10 , 0.9994065803274141 )
};
\addlegendentry{$c=50$}

\end{axis}
\end{tikzpicture}
	\caption{Probability $p$ of selecting one or more of the corrupted nodes $c$ for the DNN with layer count $l=[1, 10]$ and the number of nodes of the DDVN $n=100$}
	\label{fig:analysis_0a}
\end{figure}
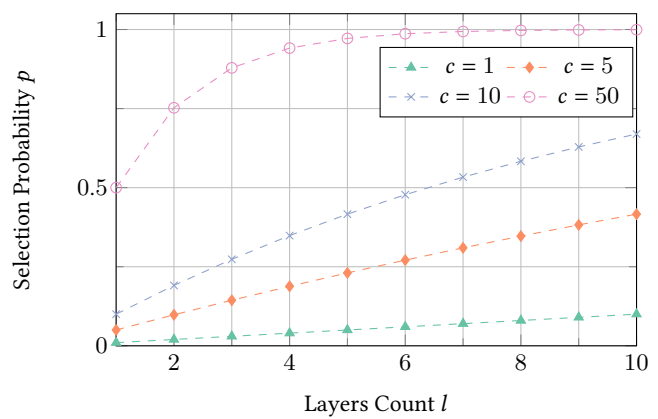

\subsubsection*{Enhancing Security with Quorums}
A quorum approach distributes the responsibility of decision-making among multiple nodes. This reduces the influence of corrupted/malicious participants since the decision is coordinated with others. Nevertheless, an attack can be successful, so below is an analysis of the advantages and disadvantages of this approach. 
Using a quorum (see  Section~\ref{sec:qudos}) implicitly increases the number of validators used. As a result, the ratio $\frac{c}{n}$ of corrupted validators $c$ to used validators $n$ decreases and percentage-wise fewer corrupted validators are found in the constructed \ac{DNN}.

\subsubsection*{The use of quorums lowers the success probability for an attack, as long as the number of corrupted nodes remains fixed} 
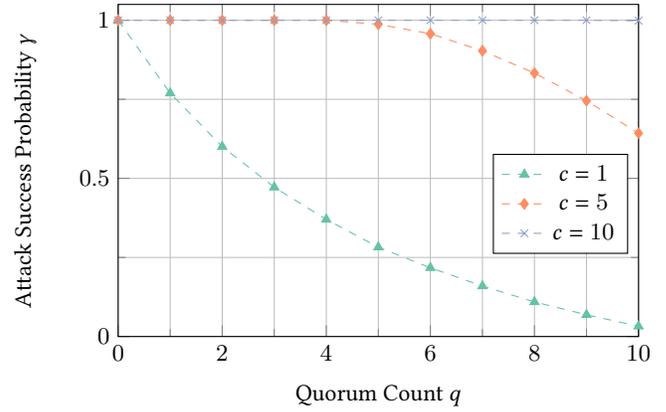
\begin{figure}[tbh]
	\centering
\begin{tikzpicture}
\begin{axis}[
height=6cm,
width=8.5cm,
grid=both, minor tick num=1,
legend style={at={(0.85,0.55)},
	anchor=north},
ylabel={Attack Success Probability $\gamma$},
xlabel={Quorum Count $q$},
      ymin=0,ymax=1.05,
      xmin=0,xmax=10
]

\addplot[dashed, mark=triangle*,color1a, mark options={solid}] coordinates {
( 0, 1.0 )
( 1, 0.76942 )
( 2, 0.60043 )
( 3, 0.47192 )
( 4, 0.37009 )
( 5, 0.2825 )
( 6, 0.21735 )
( 7, 0.15962 )
( 8, 0.10943 )
( 9, 0.06879 )
( 10, 0.03246 )
};
\addlegendentry{$c=1$}

\addplot[dashed, mark=diamond*,color2a, mark options={solid}] coordinates {
( 0, 1.0 )
( 1, 1.0 )
( 2, 1.0 )
( 3, 1.0 )
( 4, 1.0 )
( 5, 0.98721 )
( 6, 0.95726 )
( 7, 0.90343 )
( 8, 0.83291 )
( 9, 0.74555 )
( 10, 0.64286 )
};
\addlegendentry{$c=5$}

\addplot[dashed, mark=x,color3a, mark options={solid}] coordinates {
( 0, 1.0 )
( 1, 1.0 )
( 2, 1.0 )
( 3, 1.0 )
( 4, 1.0 )
( 5, 1.0 )
( 6, 1.0 )
( 7, 1.0 )
( 8, 1.0 )
( 9, 1.0 )
( 10, 0.99864 )
};
\addlegendentry{$c=10$}

\end{axis}
\end{tikzpicture}
	\caption{Comparison of the attack success probability $\gamma$ for layer count $l=10$, quorum size of $3$ and a variable quorum count $q$}
	\label{fig:analysis_1}
\end{figure}

\begin{figure}[tbh]
	\centering
\begin{tikzpicture}
\begin{axis}[
height=6cm,
width=8.5cm,
grid=both, minor tick num=1,
legend style={at={(0.85,0.98)},
	anchor=north},
ylabel={Attack Success Probability $\gamma$},
xlabel={Quorum Count $q$},
      ymin=0,ymax=1.05,
      xmin=0,xmax=10
]

\addplot[dashed, mark=triangle*,color1a, mark options={solid}] coordinates {
( 0, 1.0 )
( 1, 0.59076 )
( 2, 0.39468 )
( 3, 0.27799 )
( 4, 0.20021 )
( 5, 0.14547 )
( 6, 0.10678 )
( 7, 0.07539 )
( 8, 0.05093 )
( 9, 0.03084 )
( 10, 0.01392 )
};
\addlegendentry{$c=1$}

\addplot[dashed, mark=diamond*,color2a, mark options={solid}] coordinates {
( 0, 1.0 )
( 1, 1.0 )
( 2, 0.95579 )
( 3, 0.83893 )
( 4, 0.70493 )
( 5, 0.57207 )
( 6, 0.44969 )
( 7, 0.3414 )
( 8, 0.23913 )
( 9, 0.15186 )
( 10, 0.07354 )
};
\addlegendentry{$c=5$}

\addplot[dashed, mark=x,color3a, mark options={solid}] coordinates {
( 0, 1.0 )
( 1, 1.0 )
( 2, 1.0 )
( 3, 1.0 )
( 4, 0.97507 )
( 5, 0.90245 )
( 6, 0.7846 )
( 7, 0.65197 )
( 8, 0.49905 )
( 9, 0.34571 )
( 10, 0.19328 )
};
\addlegendentry{$c=10$}

\end{axis}
\end{tikzpicture}
	\caption{Comparison of the attack success probability $\gamma$ for layer count $l=10$, quorum size of $7$ and a variable quorum count $q$}
	\label{fig:analysis_2}
\end{figure}
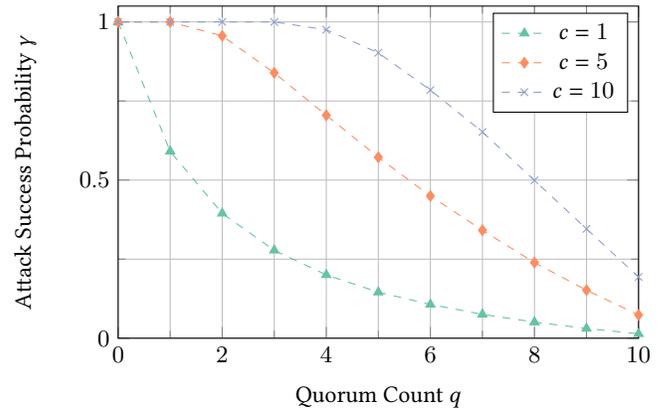

If no quorum is used and a single layer of the DNN has been corrupted, then the attacker's success probability $\gamma=1$ of the DNN. When a quorum is used, an attacker may corrupt a node in the quorum. If the number of corrupted nodes is below a certain threshold, the attacker cannot make an independent decision. Therefore, using quorums for a DNN decreases the success probability of an attack. Figure~\ref{fig:analysis_1} and Figure~\ref{fig:analysis_2} show the impact on using quorums for a DNN with a layer count $l$ of $10$. The quorum count $q$ defines how many layers are secured by a quorum. When no quorum is used, the success probability of an attack is $1$. For both figures, three different corrupted layer counts $c={1, 5, 10}$ are taken into account and $100,000$ iterations per quorum count and corrupted layer count.  

\subsubsection*{A single corrupted node has no chance of success in a DNN where each layer is protected with a quorum} The success probability for an attack in which only a single node is corrupted is $0$, as long as each layer is protected by a quorum of at least $3$ participants. This means that an attacker cannot corrupt more than $\frac{1}{2}$ of the participants, to make an independent decision. This is immediately inferred from Figure~\ref{fig:analysis_1} and Figure~\ref{fig:analysis_2} at quorum count $q = 0$.

\subsubsection*{The use of a quorum will have an adverse effect on security if the number of corrupted nodes increases proportionally to the layer size} 
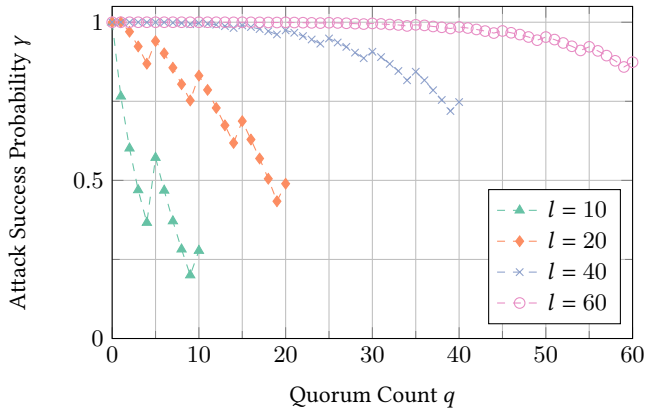
\begin{figure}[tbh]
	\centering
\begin{tikzpicture}
\begin{axis}[
height=6cm,
width=8.5cm,
grid=both, minor tick num=1,
legend style={at={(0.85,0.45)},
	anchor=north},
ylabel={Attack Success Probability $\gamma$},
xlabel={Quorum Count $q$},
      ymin=0,ymax=1.05,
      xmin=0,xmax=60
]

\addplot[dashed, mark=triangle*,color1a, mark options={solid}] coordinates {
( 0, 1.0 )
( 1, 0.76558 )
( 2, 0.60131 )
( 3, 0.46944 )
( 4, 0.36632 )
( 5, 0.57169 )
( 6, 0.46803 )
( 7, 0.37114 )
( 8, 0.28203 )
( 9, 0.20027 )
( 10, 0.27773 )
};
\addlegendentry{$l=10$}

\addplot[dashed, mark=diamond*,color2a, mark options={solid}] coordinates {
( 0, 1.0 )
( 1, 1.0 )
( 2, 0.96992 )
( 3, 0.92341 )
( 4, 0.86877 )
( 5, 0.94035 )
( 6, 0.90172 )
( 7, 0.85623 )
( 8, 0.80411 )
( 9, 0.75256 )
( 10, 0.83109 )
( 11, 0.78597 )
( 12, 0.72899 )
( 13, 0.67405 )
( 14, 0.61821 )
( 15, 0.68732 )
( 16, 0.62915 )
( 17, 0.56915 )
( 18, 0.50501 )
( 19, 0.43393 )
( 20, 0.48945 )
};
\addlegendentry{$l=20$}

\addplot[dashed, mark=x,color3a, mark options={solid}] coordinates {
( 0, 1.0 )
( 1, 1.0 )
( 2, 1.0 )
( 3, 1.0 )
( 4, 0.99956 )
( 5, 0.99994 )
( 6, 0.99952 )
( 7, 0.99854 )
( 8, 0.99695 )
( 9, 0.99377 )
( 10, 0.99712 )
( 11, 0.9951 )
( 12, 0.99231 )
( 13, 0.98752 )
( 14, 0.9818 )
( 15, 0.98984 )
( 16, 0.98475 )
( 17, 0.97844 )
( 18, 0.97157 )
( 19, 0.96138 )
( 20, 0.97423 )
( 21, 0.96673 )
( 22, 0.95617 )
( 23, 0.94542 )
( 24, 0.9322 )
( 25, 0.94865 )
( 26, 0.93597 )
( 27, 0.92204 )
( 28, 0.90391 )
( 29, 0.8863 )
( 30, 0.90664 )
( 31, 0.88978 )
( 32, 0.86832 )
( 33, 0.84581 )
( 34, 0.81644 )
( 35, 0.84359 )
( 36, 0.81652 )
( 37, 0.78502 )
( 38, 0.7542 )
( 39, 0.71923 )
( 40, 0.74764 )
};
\addlegendentry{$l=40$}

\addplot[dashed, mark=o,color4a, mark options={solid}] coordinates {
( 0, 1.0 )
( 1, 1.0 )
( 2, 1.0 )
( 3, 1.0 )
( 4, 1.0 )
( 5, 1.0 )
( 6, 1.0 )
( 7, 1.0 )
( 8, 0.99999 )
( 9, 0.99998 )
( 10, 1.0 )
( 11, 0.99999 )
( 12, 0.99994 )
( 13, 0.99984 )
( 14, 0.99967 )
( 15, 0.99992 )
( 16, 0.99971 )
( 17, 0.99954 )
( 18, 0.99928 )
( 19, 0.99911 )
( 20, 0.99963 )
( 21, 0.99904 )
( 22, 0.99857 )
( 23, 0.99801 )
( 24, 0.9973 )
( 25, 0.99841 )
( 26, 0.99775 )
( 27, 0.99669 )
( 28, 0.99556 )
( 29, 0.99419 )
( 30, 0.99624 )
( 31, 0.99466 )
( 32, 0.99284 )
( 33, 0.99143 )
( 34, 0.98825 )
( 35, 0.99185 )
( 36, 0.98999 )
( 37, 0.98658 )
( 38, 0.98401 )
( 39, 0.98027 )
( 40, 0.98502 )
( 41, 0.98167 )
( 42, 0.97714 )
( 43, 0.97199 )
( 44, 0.96574 )
( 45, 0.97239 )
( 46, 0.96665 )
( 47, 0.96095 )
( 48, 0.95333 )
( 49, 0.94323 )
( 50, 0.95307 )
( 51, 0.94516 )
( 52, 0.9344 )
( 53, 0.92395 )
( 54, 0.91082 )
( 55, 0.92226 )
( 56, 0.90942 )
( 57, 0.89447 )
( 58, 0.87804 )
( 59, 0.85852 )
( 60, 0.87376 )
};
\addlegendentry{$l=60$}

\end{axis}
\end{tikzpicture}
	\caption{Comparison of the attack success probability $\gamma$ for layer count $l=[10, 20, 40, 60]$, quorum size $3$, quorum count $q$ in the range $[0, l]$ and number of corrupted nodes $c=\frac{n}{10}$ }
	\label{fig:analysis_3}
\end{figure}

Increasing the number of corrupted nodes always results in increasing the success probability of an attack. With a proportional growth between the number of corrupted nodes and the number of layers or total nodes (the additional quorum nodes must not be neglected), with a fixed quorum size, the probability of the attacker corrupting a quorum increases (refer Figure~\ref{fig:analysis_3}). Four different DNNs are assumed, each with a layer number $l=[10, 20, 40, 60]$, a total node number of nodes $n=[30, 60, 120, 180]$ and a proportional corrupted node number of $\frac{n}{3}$. It can be seen that the attack success probability is lowest for $l=10$ and highest for $l=60$. This effect occurs when the quorum size is small compared to the number of corrupted nodes. As the corrupted number of nodes increases, but not the quorum size, the probability that an attacker can take the majority decision in a quorum increases and so does the success probability of the attack. 

\subsubsection*{Increasing the quorum size is more advantageous than increasing the layer size for enhancing QUDOS security} 

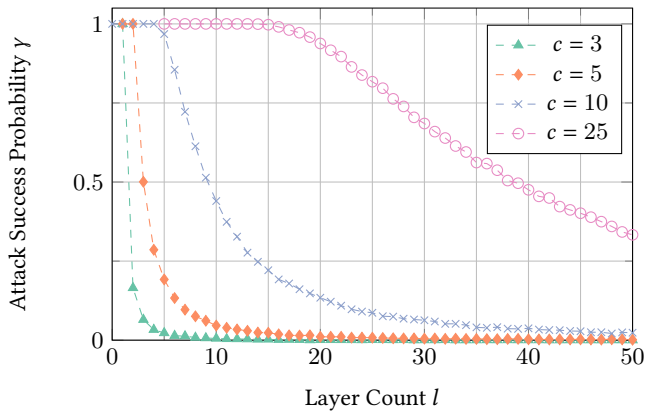
\begin{figure}[tbh]
	\centering
\begin{tikzpicture}
\begin{axis}[
height=6cm,
width=8.5cm,
grid=both, minor tick num=1,
legend style={at={(0.85,0.95)},
	anchor=north},
ylabel={Attack Success Probability $\gamma$},
xlabel={Layer Count $l$},
      ymin=0,ymax=1.05,
      xmin=0,xmax=50
]

\addplot[dashed, mark=triangle*,color1a, mark options={solid}] coordinates {
( 1, 1.0 )
( 2, 0.1653 )
( 3, 0.0649 )
( 4, 0.0332 )
( 5, 0.0228 )
( 6, 0.0141 )
( 7, 0.0128 )
( 8, 0.0083 )
( 9, 0.0068 )
( 10, 0.0054 )
( 11, 0.0041 )
( 12, 0.0038 )
( 13, 0.0033 )
( 14, 0.0024 )
( 15, 0.0025 )
( 16, 0.0015 )
( 17, 0.0018 )
( 18, 0.0014 )
( 19, 0.0004 )
( 20, 0.001 )
( 21, 0.001 )
( 22, 0.0009 )
( 23, 0.0008 )
( 24, 0.0006 )
( 25, 0.001 )
( 26, 0.0009 )
( 27, 0.0007 )
( 28, 0.0009 )
( 29, 0.0001 )
( 30, 0.0004 )
( 31, 0.0004 )
( 32, 0.0008 )
( 33, 0.0002 )
( 34, 0.0002 )
( 35, 0.0003 )
( 36, 0.0005 )
( 37, 0.0004 )
( 38, 0.0006 )
( 39, 0.0004 )
( 40, 0.0004 )
( 41, 0.0004 )
( 42, 0.0001 )
( 43, 0.0002 )
( 44, 0.0003 )
( 45, 0.0003 )
( 46, 0.0002 )
( 47, 0.0001 )
( 48, 0.0003 )
( 49, 0.0002 )
( 50, 0.0001 )
};
\addlegendentry{$c=3$}

\addplot[dashed, mark=diamond*,color2a, mark options={solid}] coordinates {
( 1, 1.0 )
( 2, 1.0 )
( 3, 0.5006 )
( 4, 0.2858 )
( 5, 0.192 )
( 6, 0.1332 )
( 7, 0.0964 )
( 8, 0.0755 )
( 9, 0.0614 )
( 10, 0.0462 )
( 11, 0.0391 )
( 12, 0.0339 )
( 13, 0.0296 )
( 14, 0.0241 )
( 15, 0.0234 )
( 16, 0.0186 )
( 17, 0.0153 )
( 18, 0.0146 )
( 19, 0.0154 )
( 20, 0.0106 )
( 21, 0.0107 )
( 22, 0.0106 )
( 23, 0.0082 )
( 24, 0.0082 )
( 25, 0.0079 )
( 26, 0.007 )
( 27, 0.0061 )
( 28, 0.0056 )
( 29, 0.0057 )
( 30, 0.0049 )
( 31, 0.0053 )
( 32, 0.0042 )
( 33, 0.0047 )
( 34, 0.0039 )
( 35, 0.0039 )
( 36, 0.0033 )
( 37, 0.0032 )
( 38, 0.004 )
( 39, 0.0041 )
( 40, 0.0029 )
( 41, 0.0024 )
( 42, 0.0022 )
( 43, 0.0026 )
( 44, 0.0024 )
( 45, 0.0026 )
( 46, 0.0022 )
( 47, 0.0019 )
( 48, 0.003 )
( 49, 0.0025 )
( 50, 0.0013 )
};
\addlegendentry{$c=5$}

\addplot[dashed, mark=x,color3a, mark options={solid}] coordinates {
( 0, 1.0 )
( 1, 1.0 )
( 2, 1.0 )
( 3, 1.0 )
( 4, 1.0 )
( 5, 0.9678 )
( 6, 0.855 )
( 7, 0.7222 )
( 8, 0.6131 )
( 9, 0.5138 )
( 10, 0.44 )
( 11, 0.3741 )
( 12, 0.3273 )
( 13, 0.2767 )
( 14, 0.2478 )
( 15, 0.2211 )
( 16, 0.1921 )
( 17, 0.1792 )
( 18, 0.1611 )
( 19, 0.1479 )
( 20, 0.1342 )
( 21, 0.1217 )
( 22, 0.1091 )
( 23, 0.0977 )
( 24, 0.0905 )
( 25, 0.0866 )
( 26, 0.0757 )
( 27, 0.0745 )
( 28, 0.068 )
( 29, 0.0654 )
( 30, 0.063 )
( 31, 0.059 )
( 32, 0.0516 )
( 33, 0.0518 )
( 34, 0.0474 )
( 35, 0.0405 )
( 36, 0.039 )
( 37, 0.0411 )
( 38, 0.0364 )
( 39, 0.0357 )
( 40, 0.0375 )
( 41, 0.0336 )
( 42, 0.0321 )
( 43, 0.0315 )
( 44, 0.0291 )
( 45, 0.0283 )
( 46, 0.025 )
( 47, 0.0251 )
( 48, 0.0211 )
( 49, 0.0246 )
( 50, 0.0236 )
};
\addlegendentry{$c=10$}

\addplot[dashed, mark=o,color4a, mark options={solid}] coordinates {
( 5, 1.0 )
( 6, 1.0 )
( 7, 1.0 )
( 8, 1.0 )
( 9, 1.0 )
( 10, 1.0 )
( 11, 1.0 )
( 12, 1.0 )
( 13, 0.9999 )
( 14, 0.9993 )
( 15, 0.9974 )
( 16, 0.9906 )
( 17, 0.9806 )
( 18, 0.9721 )
( 19, 0.9582 )
( 20, 0.9381 )
( 21, 0.9166 )
( 22, 0.8966 )
( 23, 0.864 )
( 24, 0.8382 )
( 25, 0.8172 )
( 26, 0.7967 )
( 27, 0.763 )
( 28, 0.739 )
( 29, 0.7045 )
( 30, 0.685 )
( 31, 0.6594 )
( 32, 0.6391 )
( 33, 0.6139 )
( 34, 0.5949 )
( 35, 0.5616 )
( 36, 0.5583 )
( 37, 0.5374 )
( 38, 0.5047 )
( 39, 0.4963 )
( 40, 0.4754 )
( 41, 0.4547 )
( 42, 0.4495 )
( 43, 0.4223 )
( 44, 0.4119 )
( 45, 0.4014 )
( 46, 0.3889 )
( 47, 0.3747 )
( 48, 0.3596 )
( 49, 0.3422 )
( 50, 0.3331 )
};
\addlegendentry{$c=25$}

\end{axis}
\end{tikzpicture}
	\caption{Comparison of the attack success probability $\gamma$ for layer count $l=[0, 50]$, quorum size $5$, quorum count $q=l$ and number of corrupted nodes $c$}
	\label{fig:analysis_4}
\end{figure}

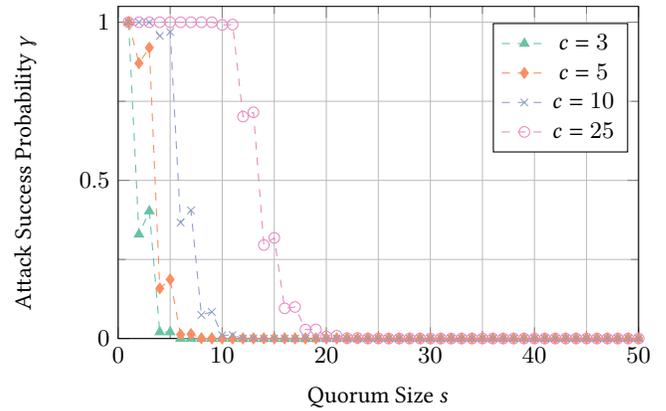
\begin{figure}[tbh]
	\centering
\begin{tikzpicture}
\begin{axis}[
height=6cm,
width=8.5cm,
grid=both, minor tick num=1,
legend style={at={(0.85,0.95)},
	anchor=north},
ylabel={Attack Success Probability $\gamma$},
xlabel={Quorum Size $s$},
      ymin=0,ymax=1.05,
      xmin=0,xmax=50
]

\addplot[dashed, mark=triangle*,color1a, mark options={solid}] coordinates {
( 1, 1.0 )
( 2, 0.3292 )
( 3, 0.4029 )
( 4, 0.0204 )
( 5, 0.0204 )
( 6, 0.0 )
( 7, 0.0 )
( 8, 0.0 )
( 9, 0.0 )
( 10, 0.0 )
( 11, 0.0 )
( 12, 0.0 )
( 13, 0.0 )
( 14, 0.0 )
( 15, 0.0 )
( 16, 0.0 )
( 17, 0.0 )
( 18, 0.0 )
( 19, 0.0 )
( 20, 0.0 )
( 21, 0.0 )
( 22, 0.0 )
( 23, 0.0 )
( 24, 0.0 )
( 25, 0.0 )
( 26, 0.0 )
( 27, 0.0 )
( 28, 0.0 )
( 29, 0.0 )
( 30, 0.0 )
( 31, 0.0 )
( 32, 0.0 )
( 33, 0.0 )
( 34, 0.0 )
( 35, 0.0 )
( 36, 0.0 )
( 37, 0.0 )
( 38, 0.0 )
( 39, 0.0 )
( 40, 0.0 )
( 41, 0.0 )
( 42, 0.0 )
( 43, 0.0 )
( 44, 0.0 )
( 45, 0.0 )
( 46, 0.0 )
( 47, 0.0 )
( 48, 0.0 )
( 49, 0.0 )
( 50, 0.0 )
};
\addlegendentry{$c=3$}

\addplot[dashed, mark=diamond*,color2a, mark options={solid}] coordinates {
( 1, 1.0 )
( 2, 0.8703 )
( 3, 0.9196 )
( 4, 0.1584 )
( 5, 0.1867 )
( 6, 0.0122 )
( 7, 0.0143 )
( 8, 0.0005 )
( 9, 0.0004 )
( 10, 0.0 )
( 11, 0.0 )
( 12, 0.0 )
( 13, 0.0 )
( 14, 0.0 )
( 15, 0.0 )
( 16, 0.0 )
( 17, 0.0 )
( 18, 0.0 )
( 19, 0.0 )
( 20, 0.0 )
( 21, 0.0 )
( 22, 0.0 )
( 23, 0.0 )
( 24, 0.0 )
( 25, 0.0 )
( 26, 0.0 )
( 27, 0.0 )
( 28, 0.0 )
( 29, 0.0 )
( 30, 0.0 )
( 31, 0.0 )
( 32, 0.0 )
( 33, 0.0 )
( 34, 0.0 )
( 35, 0.0 )
( 36, 0.0 )
( 37, 0.0 )
( 38, 0.0 )
( 39, 0.0 )
( 40, 0.0 )
( 41, 0.0 )
( 42, 0.0 )
( 43, 0.0 )
( 44, 0.0 )
( 45, 0.0 )
( 46, 0.0 )
( 47, 0.0 )
( 48, 0.0 )
( 49, 0.0 )
( 50, 0.0 )
};
\addlegendentry{$c=5$}

\addplot[dashed, mark=x,color3a, mark options={solid}] coordinates {
( 1, 1.0 )
( 2, 1.0 )
( 3, 1.0 )
( 4, 0.9567 )
( 5, 0.9694 )
( 6, 0.3667 )
( 7, 0.4051 )
( 8, 0.0744 )
( 9, 0.0841 )
( 10, 0.0102 )
( 11, 0.011 )
( 12, 0.0002 )
( 13, 0.0007 )
( 14, 0.0001 )
( 15, 0.0002 )
( 16, 0.0 )
( 17, 0.0 )
( 18, 0.0 )
( 19, 0.0 )
( 20, 0.0 )
( 21, 0.0 )
( 22, 0.0 )
( 23, 0.0 )
( 24, 0.0 )
( 25, 0.0 )
( 26, 0.0 )
( 27, 0.0 )
( 28, 0.0 )
( 29, 0.0 )
( 30, 0.0 )
( 31, 0.0 )
( 32, 0.0 )
( 33, 0.0 )
( 34, 0.0 )
( 35, 0.0 )
( 36, 0.0 )
( 37, 0.0 )
( 38, 0.0 )
( 39, 0.0 )
( 40, 0.0 )
( 41, 0.0 )
( 42, 0.0 )
( 43, 0.0 )
( 44, 0.0 )
( 45, 0.0 )
( 46, 0.0 )
( 47, 0.0 )
( 48, 0.0 )
( 49, 0.0 )
( 50, 0.0 )
};
\addlegendentry{$c=10$}

\addplot[dashed, mark=o,color4a, mark options={solid}] coordinates {
( 1, 1.0 )
( 2, 1.0 )
( 3, 1.0 )
( 4, 1.0 )
( 5, 1.0 )
( 6, 1.0 )
( 7, 1.0 )
( 8, 1.0 )
( 9, 1.0 )
( 10, 0.9916 )
( 11, 0.9925 )
( 12, 0.7015 )
( 13, 0.7152 )
( 14, 0.2957 )
( 15, 0.3182 )
( 16, 0.0954 )
( 17, 0.0996 )
( 18, 0.0283 )
( 19, 0.0289 )
( 20, 0.0064 )
( 21, 0.0088 )
( 22, 0.0011 )
( 23, 0.0017 )
( 24, 0.0001 )
( 25, 0.0003 )
( 26, 0.0 )
( 27, 0.0 )
( 28, 0.0 )
( 29, 0.0 )
( 30, 0.0 )
( 31, 0.0 )
( 32, 0.0 )
( 33, 0.0 )
( 34, 0.0 )
( 35, 0.0 )
( 36, 0.0 )
( 37, 0.0 )
( 38, 0.0 )
( 39, 0.0 )
( 40, 0.0 )
( 41, 0.0 )
( 42, 0.0 )
( 43, 0.0 )
( 44, 0.0 )
( 45, 0.0 )
( 46, 0.0 )
( 47, 0.0 )
( 48, 0.0 )
( 49, 0.0 )
( 50, 0.0 )
};
\addlegendentry{$c=25$}

\end{axis}
\end{tikzpicture}
	\caption{Comparison of the attack success probability $\gamma$ for quorum size $s=[0, 50]$, layer count $l=5$, quorum count $q=l$ and number of corrupted nodes $c$}
	\label{fig:analysis_5}
\end{figure}

By increasing the participants in the quorum, it becomes more difficult for an attacker to corrupt a quorum. Figure~\ref{fig:analysis_4} and Figure~\ref{fig:analysis_5} compare the two approaches of increasing the number of layers and increasing the quorum size. It can be seen that for a total node count of $n=l \dot s=75$, increasing the quorum size drastically decreases the success probability of an attack. For $c=10$, where $\gamma_l > 0.22$ (increasing layers) and $\gamma_s<0.01$ (increasing quorum size), the improvement is more than $20\%$. Examining $c=25$, the difference between increasing the layer count and increasing the quorum count is even greater $\gamma_l>0.99$ and $\gamma_s<0.32$, which is more than $67\%$.

In the range of $l=[1, 5]$ and respectively $s=[1, 5]$, demonstrates superior performance at first glance. However, due to the assumed fixed quorum size $s=5$, the layer approach has a larger quorum size and performs better in this range. When the quorum size $s=5$ is exceeded, the approaches perform as expected. 

\subsubsection*{Increasing the threshold value for a quorum decision decreases the success probability of an attack.} 
\begin{figure}[hbt]
	\centering
\begin{tikzpicture}
\begin{axis}[
height=6cm,
width=8.5cm,
grid=both, minor tick num=1,
legend style={at={(0.85,0.95)},
	anchor=north},
ylabel={Attack Success Probability $\gamma$},
xlabel={Convinced Nodes $n_{min}$},
      ymin=0,ymax=1.05,
      xmin=6,xmax=11
]

\addplot[dashed, mark=triangle*,color1a, mark options={solid}] coordinates {
( 6, 0.0442 )
( 7, 0.005 )
( 8, 0.0 )
( 9, 0.0 )
( 10, 0.0 )
( 11, 0.0 )
};
\addlegendentry{$c=20$}

\addplot[dashed, mark=diamond*,color2a, mark options={solid}] coordinates {
( 6, 0.9211 )
( 7, 0.3071 )
( 8, 0.0719 )
( 9, 0.011 )
( 10, 0.0011 )
( 11, 0.0 )
};
\addlegendentry{$c=40$}

\addplot[dashed, mark=x,color3a, mark options={solid}] coordinates {
( 6, 1.0 )
( 7, 0.6242 )
( 8, 0.2896 )
( 9, 0.0963 )
( 10, 0.0193 )
( 11, 0.0021 )
};
\addlegendentry{$c=60$}

\addplot[dashed, mark=o,color4a, mark options={solid}] coordinates {
( 6, 1.0 )
( 7, 0.8932 )
( 8, 0.6837 )
( 9, 0.3977 )
( 10, 0.1448 )
( 11, 0.0258 )
};
\addlegendentry{$c=80$}

\end{axis}
\end{tikzpicture}
	\caption{Comparison of the attack success probability $\gamma$ for quorum size $s=11$, layer count $l=10$, quorum count $q=l$ and number of corrupted nodes $c$}
	\label{fig:analysis_6}
\end{figure}
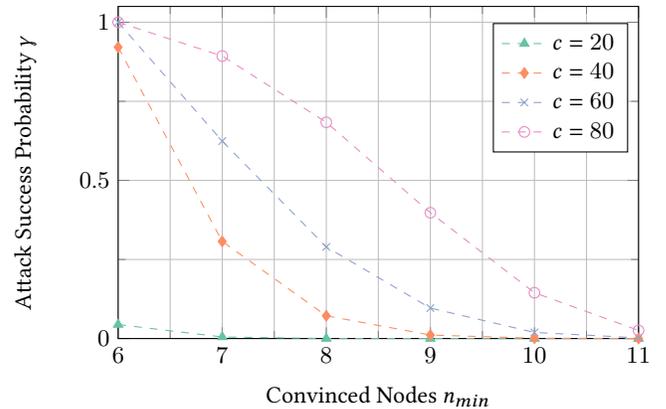

For decision-making in a quorum, the majority must agree on a decision. The number of matching decisions needed for the majority decision is called  $n_{min}$ (see Section~\ref{sec:qudos}). With an increasing number of participants in a quorum, more participants must match in their decision. This observation is utilized by \texttt{QUDOS} to reduce the probability of success of an attacker since they need to corrupt at least a defined threshold. In Figure~\ref{fig:analysis_6}, a DNN with $l=10$ and a quorum size $s=11$ is given. It is shown how likely it is for the attacker to convince at least one of the quorums of their result, i.e. to corrupt sufficient nodes. With $c=20$, the attacker has only a low probability of success $\gamma<0.1\%$ even with $n_{min}=6$. With a very high proportion of corrupted nodes $\frac{8}{11}$, the attacker's probability of success is $\gamma<0.1\%$ assuming that the attacker must achieve a majority in at least one quorum. 

\subsubsection*{Summary}
The following are the key observations from the above experimental analysis:

\begin{itemize}
    \item By using the \texttt{QUDOS} approach, the success probability of an attack on DNN layers is reduced. In the best case, with only one corrupted node, the attack is completely prevented.
    \item When setting up \texttt{QUDOS}, a large number of quorum participants should be preferred to a high number of layers. If the number of layers cannot be influenced due to a given DNN model the security will be improved by increasing the quorum size.
    \item By increasing the threshold $t$, the success probability for an attack can be greatly reduced since an attack must always exceed the threshold of corrupted nodes to make a decision in a quorum. 
\end{itemize}
\section{Conclusion and Outlook}
\label{sec:outlook}
This paper presents a quorum-based security enhancement mechanism for DNNs, called \texttt{QUDOS}. \texttt{QUDOS} combines the DDVN's capability to solve the single point of failure with the ability to protect sequential algorithms, for example DNNs, by a quorum. Due to the sequential coupling of layers, DNNs have the problem that a single corrupted layer is sufficient to manipulate the overall prediction result. \texttt{QUDOS} protects against this by replicating a single layer on several nodes. The result of a quorum respectively a layer corresponds to the majority decision of the individual nodes. This means that an attacker has always to provide the majority in a quorum so that they can make the decision. For the analysis of \texttt{QUDOS}, the corruption factor and the success probability for an attack on a DNN were defined. In addition, a formula for the selection probability of corrupted nodes when using a DDVN was given. 

The analysis contains a comprehensive analysis describing different attacker knowledge a) oblivious, b) expected result of the layer, c) expected result of the DNN, d) input data of the DNN and e) DNN structure. For the analysis, an oblivious attacker knowledge is assumed and based on that, it is shown that:

\begin{itemize}
    \item By using a DDVN, the selection probability of corrupted nodes decreases and thus the security of the DNN increases.
    \item Attacks with a low number of corrupted nodes can be completely prevented or its impact reduced by \texttt{QUDOS}.
    \item Increasing the quorum size is better in terms of security than increasing the number of layers.
    \item Increasing the threshold in a quorum has a positive effect on security.
\end{itemize}

\subsubsection*{Limitations and Future Work} 
\texttt{QUDOS} assumes that all nodes of a quorum are equally trustworthy. In real scenarios, this may not be the case. Therefore, approaches that consider trust levels and make a distinction between the nodes of a quorum and their influence on the final decision are required. Efforts will be made towards comparing security analysis with other data structures to determine the most appropriate data structure for a given use case. Explorations will be undertaken that consider the effect of DNNs with a large number of layers (greater than $500$) and how the results are obtained.

\balance
\bibliography{main}

\end{document}